\documentclass[]{piparticle-final}
\usepackage{graphicx}
\usepackage{amsmath}
\usepackage{ulem}  

\begin{document}

\volume{1}               
\articlenumber{010001}   
\journalyear{2009}       
\editor{S. A. Cannas}   
\received{17 June 2009}     
\accepted{28 August 2009}   
\runningauthor{S. A. Sartarelli \itshape{et al.}}  
\doi{010001}         

\title{Correlation between asymmetric profiles in slits and standard
prewetting lines}

\author{Salvador A. Sartarelli,\cite{inst1}\thanks{E-mail: asarta@ungs.edu.ar}\hspace{2mm}
        Leszek Szybisz\cite{inst2}$^-$\cite{inst4}\thanks{E-mail: szybisz@tandar.cnea.gov.ar}} 

\pipabstract{The adsorption of Ar on substrates of Li is investigated
within the framework of a density functional theory which includes an
effective pair potential recently proposed. This approach yields good
results for the surface tension of the liquid-vapor interface over the 
entire range of temperatures, $T$, from the triple point, $T_t$, to 
the critical point, $T_c$. The behavior of the adsorbate in the cases
of a single planar wall and a slit geometry is analyzed as a function
of temperature. Asymmetric density profiles are found for fluid
confined in a slit built up of two identical planar walls leading to
the spontaneous symmetry breaking (SSB) effect. We found that the
asymmetric solutions occur even above the wetting temperature $T_w$ in
a range of average densities $\rho^*_{ssb1} \le \rho^*_{av} \le
\rho^*_{ssb2}$, which diminishes with increasing temperatures until
its disappearance at the critical prewetting point $T_{cpw}$. In this
way a correlation between the disappearance of the SSB effect and the
end of prewetting lines observed in the adsorption on a one-wall planar
substrate is established. In addition, it is shown that a value for
$T_{cpw}$ can be precisely determined by analyzing the asymmetry
coefficients.
}

\maketitle

\blfootnote{
\begin{theaffiliation}{99}
   \institution{inst1} Instituto de Desarrollo Humano, Universidad
Nacional de General Sarmiento, Gutierrez 1150, RA--1663 San Miguel, Argentina.
   \institution{inst2} Laboratorio TANDAR, Departamento de F\'{\i}sica, Comisi\'on Nacional de Energ\'{\i}a At\'omica, Av. del Libertador 8250, RA--1429 Buenos Aires, Argentina.
   \institution{inst3} Departamento de F\'{\i}sica, Facultad de Ciencias Exactas y Naturales, Universidad de Buenos Aires, Ciudad Universitaria, RA--1428 Buenos Aires, Argentina.
   \institution{inst4} Consejo Nacional de Investigaciones Cient\'{\i}ficas y T\'ecnicas, Av. Rivadavia 1917, RA--1033 Buenos Aires, Argentina.

\end{theaffiliation}
}

\section{Introduction}

The study of physisorption of fluids on solid substrates had led to
very fascinating phenomena mainly determined by the relative strengths
of fluid-fluid ($f$-$f$) and substrate-fluid ($s$-$f$) attractions.
In the present work we shall refer to two of such features. One is the
prewetting curve identified in the study of fluids adsorbed on planar
surfaces above the wetting temperature $T_w$ (see, e.g., Pandit, Schick,
and Wortis \cite{pandit82}) and the other is the occurrence of
asymmetric profiles of fluids confined in a slit of identical walls
found by van Leeuwen and collaborators in molecular dynamics
calculations \cite{sikkenk87,nijmeijer90}. It is known that for a
strong substrate (i.e., when the $s$-$f$ attraction dominates over the
$f$-$f$ one) the adsorbed film builds up continuously showing a
complete wetting.In such a case, neither prewetting transitions nor
spontaneous symmetry breaking (SSB) of the profiles are observed, both
these phenomena appear for substrates of moderate strength.

The prewetting has been widely analyzed for adsorption of quantum as
well as classical fluids. A summary of experimental data and theoretical
calculations for $^4$He may be found in Ref.\ \cite{szybisz03}. Studies
of other fluids are mentioned in Ref.\ \cite{sartarelli09}. These
investigations indicated that prewetting is present in real systems
such as $^4$He, H$_2$, and inert gases adsorbed on alkali metals.

On the other hand, after a recent work of Berim and Ruckenstein
\cite{berim07} there is a renewal of the interest in searching for the
SSB effect in real systems. These authors utilized a density functional
(DF) theory to study the confinement of Ar in a slit composed of two
identical walls of CO$_2$ and concluded that SSB occurs in a certain
domain of temperatures. In a revised analysis of this case, reported in
Ref.\ \cite{szybisz08}, we found that the conditions for the SSB were
fulfilled because the authors of Ref.\ \cite{berim07} had diminished
the $s$-$f$ attraction by locating an extra hard-wall repulsion.
However, it was found that inert gases adsorbed on alkali metals
exhibit SSB. Results for Ne confined by such substrates were recently
reported \cite{sartarelli09b}.

The aim of the present investigation is to study the relation between
the range of temperatures where the SSB occurs and the temperature
dependence of the wetting properties. In this paper we illustrate
our findings describing the results for Ar adsorbed on Li. Previous DF
calculations of Ancilotto and Toigo \cite{ancilotto99} as well as Grand
Canonical Monte Carlo (GCMC) simulations carried out by Curtarolo {\it
et al.} \cite{curtarolo00} suggest that Ar wets Li at a temperature
significantly below $T_c$. So, this system should exhibit a large locus
of the prewetting line and this feature makes it very convenient for
our study as it was already communicated during a recent workshop
\cite{szybisz09}.

The paper is organized in the following way. The theoretical background
is summarized in Sec.\ \ref{sec:theory}. The results, together with
their analysis, are given in Sec.\ \ref{sec:results}. Sec.\
\ref{sec:conclusions} is devoted to the conclusions.

\section{Theoretical background}
\label{sec:theory}

In a DF theory, the Helmholtz free energy $F_{\rm DF}[\rho({\bf r})]$
of an inhomogeneous fluid embedded in an external potential
$U_{sf}({\bf r})$ is expressed as a functional of the local density
$\rho({\bf r})$ (see, e.g.,  Ref.\ \cite{ravikovitch01})

\begin{eqnarray}
&&F_{\rm DF}[\rho({\bf r})] \nonumber\\
&=& \nu_{\rm id}\,k_B\,T \int d{\bf r}\,
\rho({\bf r})\,\{\ln[\Lambda^3\rho({\bf r})]-1\} \nonumber\\
&+& \int d{\bf r}\,\rho({\bf r})\,f_{\rm HS}[\bar\rho({\bf r});
d_{\rm HS}] \nonumber\\
&+& \frac{1}{2} \int \int d{\bf r}\,d{\bf r\prime}\,\rho({\bf r})\,
\rho({\bf r\prime\prime})\,\Phi_{\rm attr}(\mid {\bf r}-{\bf r\prime}
\mid ) \nonumber\\
&+& \int d{\bf r}\,\rho({\bf r})\,U_{sf}({\bf r}) \;. \label{free0}
\end{eqnarray}
The first term is the ideal gas free energy, where $k_B$ is the
Boltzmann constant and $\Lambda=\sqrt{2\,\pi\,\hbar^2/m\,k_B\,T}$ the
de Broglie thermal wavelength of the molecule of mass $m$. Quantity
$\nu_{\rm id}$ is a parameter introduced in Eq.\ (2) of
\cite{ancilotto01} (in the standard theory it is equal unity). The
second term accounts for the repulsive $f$-$f$ interaction
approximated by a hard-sphere (HS) functional with a certain choice
for the HS diameter $d_{\rm HS}$. In the present work we have used for
$f_{\rm HS}[\bar\rho({\bf r});d_{\rm HS}]$ the expression provided by
the nonlocal DF (NLDF) formalism developed by Kierlik and Rosinberg
\cite{kierlik90} (KR), where $\bar\rho({\bf r})$ is a properly
averaged density. The third term is the attractive $f$-$f$ interactions
treated in a mean field approximation (MFA). Finally, the last integral
represents the effect of the external potential $U_{sf}({\bf r})$
exerted on the fluid.

In the present work, for the analysis of physisorption we adopted the
{\it ab initio} potential of Chismeshya, Cole, and Zaremba (CCZ)
\cite{ccz98} with the parameters listed in Table 1 therein.

\subsection{Effective pair attraction}
\label{sec:pair}

The attractive part of the $f$-$f$ interaction was described by an
effective pair interaction devised in Ref.\ \cite{sartarelli09}, where
the separation of the Lennard-Jones (LJ) potential introduced by
Weeks, Chandler and Andersen (WCA) \cite{weeks70} is adopted

\begin{eqnarray}
&&\Phi^{\rm WCA}_{\rm attr}(r) \nonumber\\
&=& \left\{ \begin{array}{ll}
   - \tilde{\varepsilon}_{ff} \;, & r \le r_m \, \\
   4 \tilde{\varepsilon}_{ff} \biggr[ 
     \left(\frac{\tilde{\sigma}_{ff}}{r}\right)^{12}
     - \left(\frac{\tilde{\sigma}_{ff}}{r}\right)^6 \biggr] \;,
   & r > r_m \;.
         \end{array} \right.
\nonumber\\
\label{weeks}
\end{eqnarray}
Here $r_m=2^{1/6}\tilde{\sigma}_{ff}$ is the position of the LJ
minimum. No cutoff for the pair potential was introduced. The well
depth $\tilde{\varepsilon}_{ff}$ and the interaction size
$\tilde{\sigma}_{ff}$ are considered as free parameters because the use
of the bare values $\varepsilon_{ff}/k_B = 119.76$~K and $\sigma_{ff} =
3.405$~\AA$\,$ overestimates $T_c$.

So, the complete DF formalism has three adjustable parameters (namely,
$\nu_{id}$, $\tilde{\varepsilon}_{ff}$, and $\tilde{\sigma}_{ff}$),
which were determined by imposing that at $l$-$v$ coexistence, the
pressure as well as the chemical potential of the bulk $l$ and $v$
phases should be equal [i.e., $P(\rho_l)=P(\rho_v)$ and $\mu(\rho_l)=
\mu(\rho_v)$]. The procedure is described in Ref. \cite{sartarelli09}.
In practice, we set $d_{\rm HS}=\tilde{\sigma}_{ff}$ and imposed 
the coexistence data of $\rho_l$, $\rho_v$, and $P(\rho_l)=P(\rho_v)=
P_0$ for Ar quoted in Table~X of Ref.\ \cite{rabinovich88} to be 
reproduced in the entire range of temperatures $T$ between $T_t=
83.78$~K and $T_c=150.86$~K.

\subsection{Euler-Lagrange equation}
\label{sec:Euler}

The equilibrium density profile $\rho({\bf r})$ of the adsorbed fluid
is determined by a minimization of the free energy with respect to
density variations with the constraint of a fixed number of particles
$N$

\begin{equation}
\frac{\delta}{\delta \rho({\bf r})} \biggr[ F_{\rm DF}[\rho({\bf r})]
- \mu \int d{\bf r}\,\rho({\bf r}) \biggr] = 0 \;. \label{euler-l}
\end{equation}
Here the Lagrange multiplier $\mu$ is the chemical potential of the
system. In the case of a planar symmetry where the flat walls exhibit
an infinite extent in the $x$ and $y$ directions, the profile depends
only on the coordinate $z$ perpendicular to the substrate. For this
geometry, the variation of Eq.\ (\ref{euler-l}) yields the following
Euler-Lagrange (E-L) equation

\begin{eqnarray}
\frac{\delta [(F_{\rm id} + F_{\rm HS})/A]}{\delta \rho(z)}
&+& \int^L_0 dz\prime \rho(z\prime) \bar\Phi_{\rm attr}(\mid z
- z\prime \mid) \nonumber\\
&+& U_{sf}(z) = \mu \;, \label{euler}
\end{eqnarray}
where

\begin{equation}
\frac{\delta (F_{\rm id}/A)}{\delta \rho(z)}
= \nu_{\rm id}\,k_B\,T\,\ln{[\Lambda^3\,\rho(z)]} \;, \label{ideal}
\end{equation}
and 

\begin{eqnarray}
&&\frac{\delta (F_{\rm HS}/A)}{\delta \rho(z)}
= f_{\rm HS}[\bar\rho(z);d_{\rm HS}] \nonumber\\
&+& \int^L_0 dz\prime\,\rho(z\prime)
\,\frac{\delta f_{\rm HS}[\bar\rho(z\prime);d_{\rm HS}]}
{\delta \bar\rho(z\prime)}
\,\frac{\delta \bar\rho(z\prime)}{\delta \rho(z)}
\;. \label{film}
\end{eqnarray}
Here $F_{\rm id}/A$ and $F_{\rm HS}/A$ are free energies per unit of
one wall area $A$. $L$ is the size of the box adopted for solving
the E-L equations. The boundary conditions for the one-wall and slit
systems are different and will be given below. The final E-L equation
may cast into the form

\begin{eqnarray}
\nu_{\rm id}\,k_B\,T\,\ln{[\Lambda^3\,\rho(z)]} + Q(z) = \mu \;,
\label{euler1}
\end{eqnarray}
where

\begin{eqnarray}
Q(z)&=&f_{\rm HS}[\bar\rho(z);d_{\rm HS}] \nonumber\\
&+& \int^L_0 dz\prime\,\rho(z\prime)
\,\frac{\delta f_{\rm HS}[\bar\rho(z\prime);d_{\rm HS}]}
{\delta \bar\rho(z\prime)}
\,\frac{\delta \bar\rho(z\prime)}{\delta \rho(z)} \nonumber\\
&+& \int^L_0 dz\prime\,\rho(z\prime)\,
\bar\Phi_{\rm attr}(\mid z - z\prime \mid) \nonumber\\
&+& U_{sf}(z) \;. \label{euler2}
\end{eqnarray}
The number of particles $N_s$ per unit area, $A$, of the wall is

\begin{equation}
N_s = \frac{N}{A} = \int^L_0 \rho(z)\,dz \;. \label{cover}
\end{equation}

In order to get solutions for $\rho(z)$, it is useful to rewrite Eq.\
(\ref{euler1}) as

\begin{equation}
\rho(z) = \rho_0\,\exp{\left(-\frac{Q(z)}{\nu_{\rm id}\,k_B\,T}\right)}
\;, \label{rho}
\end{equation}
with

\begin{equation}
\rho_0 = \frac{1}{\Lambda^3}
\exp{\left(\frac{\mu}{\nu_{\rm id}\,k_B\,T}\right)} \;. \label{rho0}
\end{equation}
The relation between $\mu$ and $N_s$ is obtained by substituting
Eq.\ (\ref{rho}) into the constraint of Eq.\ (\ref{cover})

\begin{eqnarray}
\mu &=& - \nu_{\rm id}\,k_B\,T \nonumber\\
&\times& \ln{\biggr[\frac{1}{N_s \Lambda^3}\int^L_0 dz
\exp{\left(-\frac{Q(z)}{\nu_{\rm id} k_B T}\right) \biggr]}} \;.
\label{lambda}
\end{eqnarray}

When solving this kind of systems, it is usual to define dimensionless
variables $z^* = z/{\tilde\sigma}_{ff}$ for the distance and $\rho^* =
\rho\,{\tilde\sigma}^3_{ff}$ for the densities. In these units the box
size becomes $L^*=L/{\tilde\sigma}_{ff}$.

\section{Results and Analysis}
\label{sec:results}

In order to quantitatively study the adsorption of fluids within any
theoretical approach,one must require the experimental surface
tension of the bulk liquid-vapor interface, $\gamma_{lv}$, to be
reproduced satisfactorily over the entire $T_t \le T \le T_c$
temperature range. Therefore, we shall first examine the prediction for this observable before studying the adsorption phenomena.

\subsection{Surface tension of the bulk liquid-vapor interface}
\label{sec:tension}

\begin{figure}
\centering\includegraphics[width=0.48\textwidth]{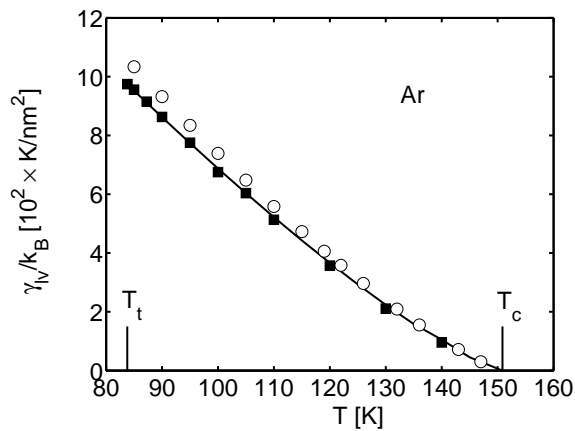}   
\caption{\label{fig:Ar_stens} Surface tension of Ar as a function of
temperature. Squares are experimental data taken from Table II of Ref.\
\cite{wuyan82}. The solid curve corresponds to the fluctuation theory
of critical phenomena and the circles are present DF results.}
\end{figure}

Figure \ref{fig:Ar_stens} shows the experimental data of $\gamma_{lv}$
taken from Table II of Ref.\ \cite{wuyan82}. In order to theoretically evaluate
this quantity the E-L equations for free slabs of Ar,
i.e. setting

\begin{equation}
U_{sf}(z)=0 \;, \label{free}
\end{equation}
were solved imposing periodic boundary conditions $\rho(z=0)=\rho(z=L)$.
At a given temperature $T$, for a sufficiently large system one must
obtain a wide central region with $\rho(z \simeq L/2)=\rho_l(T)$ and
tails with density $\rho_v(T)$, where the values of $\rho_l(T)$ and
$\rho_v(T)$ should be those of the liquid-vapor coexistence curve.   
The surface tension of the liquid-vapor interface is calculated
according to the thermodynamic definition

\begin{equation}
\gamma_{lv} = (\Omega+P_0\,V)/A = \Omega/A + P_0\,L \;,
\label{lv-tension}
\end{equation}
where $\Omega=F_{\rm DF}-\mu\,N$ is the grand potential of the system
and $P_0$ the pressure at liquid-vapor coexistence previously introduced. We solved a box with $L^*=40$. The obtained results are
plotted in Fig.\ \ref{fig:Ar_stens} together with the prediction of
the fluctuation theory of critical phenomena $\gamma_{lv}=\gamma^0_{lv}
(1-T/T_c)^{1.26}$ with $\gamma^0_{lv}=17.4$~K/\AA$^2$ (see, e.g.,
\cite{vrabec06}). One may realize that our values are in satisfactory
agreement with experimental data and the renormalization theory over the
entire range of temperatures $T_t \le T \le T_c$, showing a small
deviation near $T_t$.

\begin{figure}[th]
\centering\includegraphics[width=0.48\textwidth]{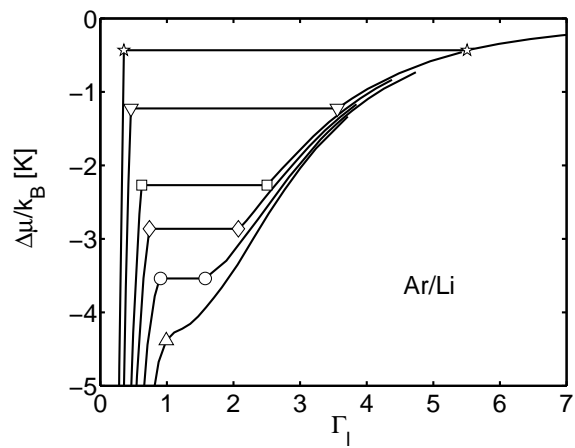}   
\caption{\label{fig:ArLi_Dmu} Adsorption isotherms for the Ar/Li
system, i.e., $\Delta\mu$ as a function of coverage $\Gamma_\ell$.
Up-triangles correspond to $T=119$~K; circles to $T=118$~K; diamonds
to $T=117$~K; squares to $T=116$~K; down-triangles to $T=114$~K and
stars to $T=112$~K.}
\end{figure}

\subsection{Adsorption on one planar wall}
\label{sec:one_wall}

\begin{figure}[th]
\centering\includegraphics[width=0.48\textwidth]{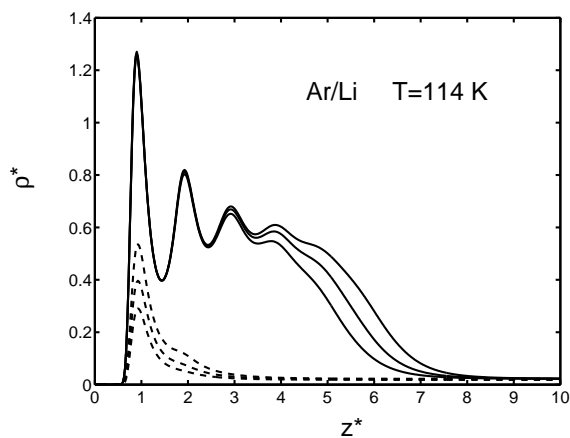}
\caption{\label{fig:ArLi_114} Examples of density profiles of Ar
adsorbed on a surface of Li at $T=114$~K displayed as a function of
the distance from the wall located at $z^*=0$. Dashed curves are
profiles for $\Gamma_\ell$ below the coverage jump, while solid curves
are stable films above this jump.}
\end{figure}

It is assumed that the physisorption of Ar on a one wall substrate of
Li is driven by the CCZ potential, i.e.,

\begin{equation}
U_{sf}(z)=U_{\rm CCZ}(z) \;. \label{wall}
\end{equation}
The E-L equations were solved in a box of size $L^*=40$ by imposing
$\rho(z>L)=\rho(z=L)$. The solution gives a density profile $\rho(z)$
and the corresponding chemical potential $\mu$. Adsorption isotherms
at a given temperature were calculated as function of the excess surface
density. This quantity, also termed coverage, is often expressed in
nominal layers $\ell$

\begin{equation}
\Gamma_\ell = (1/\rho^{2/3}_l) \int_0^\infty dz [\rho(z) - \rho_B ] \;,
\label{coverage}
\end{equation}
where $\rho_B = \rho(z \to \infty)$ is the asymptotic bulk density and
$\rho_l$ the liquid density at saturation for a given temperature. By
utilizing the results for $\mu$ obtained from the E-L equation and the
value $\mu_0$ corresponding to saturation at a given temperature $T$,
the difference $\Delta \mu = \mu - \mu_0$ was evaluated. Figure
\ref{fig:ArLi_Dmu} shows the adsorption isotherms for temperatures
above $T_w$, where an equal area Maxwell construction is feasible.
This is just the prewetting region characterized by a jump in coverage
$\Gamma_\ell$. The size of this jump depends on temperate. The largest jump occurs at $T_w$ and diminishes for increasing $T$ until its
disappearance at $T_{cpw}$. Density profiles just below and above the
coverage jump for $T=114$~K are displayed in Fig.\ \ref{fig:ArLi_114},
in that case $\Gamma_\ell$ jumps from 0.5 to 3.6. Therefore, the
formation of the fourth layer may be observed in the plot.

\begin{figure}[th]
\centering\includegraphics[width=0.48\textwidth]{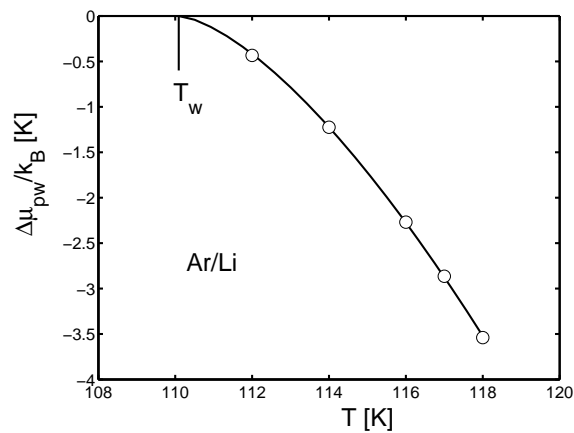}
\caption{\label{fig:PW_ArLi} Prewetting line for Ar adsorbed on Li.
The solid curve is the fit to Eq.\ (\ref{Dmu_0}) and reaches the
$\Delta\mu_{pw}/k_B=0$ line at $T_w=110.1$~K.}
\end{figure}

The wetting temperature $T_w$ can be obtained from the analysis of the
values of $\Delta\mu/k_B$ at which the jump in coverage occurs at each
considered temperature. The behavior $\Delta\mu_{pw}/k_B \, \rm vs \,T$
is displayed in Fig.\ \ref{fig:PW_ArLi}. A useful form for determining
the temperature $T_w$ was derived from thermodynamic arguments
\cite{cheng93b}

\begin{eqnarray}
\Delta \mu_{pw}(T) &=& \mu_{pw}(T) - \mu_0(T) \nonumber\\
&=& a_{pw}\,(T-T_w)^{3/2} \;.
\label{Dmu_0}
\end{eqnarray}  
Here $a_{pw}$ is a model parameter and the exponent $3/2$ is fixed by
the power of the van der Walls tail of the adsorption potential
$U_{sf}(z) \simeq -C_3/z^3$. The fit of the data of $\Delta\mu/k_B$ to
Eq.\ (\ref{Dmu_0}) yielded $T_w=110.1$~K and $a_{pw}/k_B =
-0.16$~K$^{-1/2}$.

On the other hand, according to Fig.\ \ref{fig:ArLi_Dmu}, the critical
prewetting point $T_{cpw}$ lies between $T=118$ and 119~K. At the latter
temperature, the film already presents a continuous growth. 

Our values of $T_w$ and $T_{cpw}$ are smaller than those obtained from
prior DF calculations \cite{ancilotto99} ($T_w=123$~K and $T_{cpw}
\simeq 130$~K) and GCMC simulations \cite{curtarolo00} ($T_w=130$~K).
The difference with the DF evaluation of Ref.\ \cite{ancilotto99} is
due to the use of different effective pair potentials as we explain in
Ref.\ \cite{sartarelli09}, where the adsorption of Ne is studied. The
present approach gives a reasonable $\gamma_{lv}$, while that of Ref.\
\cite{ancilotto99} fails dramatically close to $T_t$. The difference
with the GCMC results cannot be interpreted in a straightforward way.

\subsection{Confinement in a planar slit}
\label{sec:slit}

In the slit geometry, where the Ar atoms are confined by two identical
walls of Li the $s$-$f$ potential becomes

\begin{equation}
U_{sf}(z)=U_{\rm CCZ}(z) + U_{\rm CCZ}(L-z) \;. \label{slit}
\end{equation}
The walls were located at a distance $L^*=40$, this width guarantees
that the pair interaction between two atoms located at different walls
is negligible. In fact, this width is wider than $L^*=29.1$, which was
utilized in the pioneering molecular dynamics calculations
\cite{sikkenk87,nijmeijer90}. Accordingly, the E-L equations were
solved in a box of size $L^*=40$. In this geometry, the repulsion at
the walls causes the profiles $\rho(z=0)$ and $\rho(z=L)$ to be equal to
zero. The solutions were obtained at a fixed dimensionless average
density defined in terms of $N$, $A$, and $L$ as $\rho^*_{av} =
N\,{\tilde\sigma}^3_{ff}/A\,L = N^*_s/L^*$.

\begin{figure}[th]
\centering\includegraphics[width=0.48\textwidth]{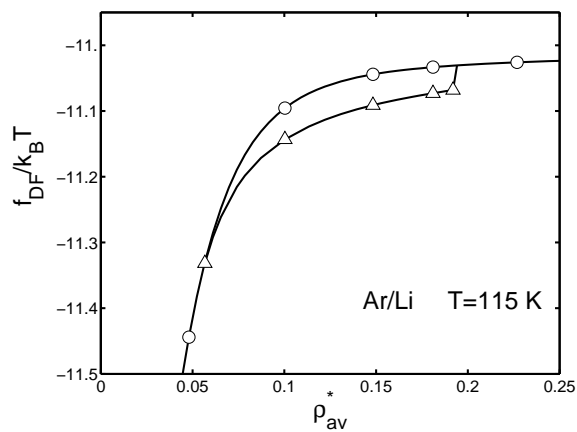}
\caption{\label{fig:free_en} Free energy per particle (in units of
$k_B\,T$) for Ar confined in a slit of Li with $L^*=40$ at $T=115$~K
displayed as a function of the average density. The curve labeled by
circles corresponds to symmetric solutions, while that labeled by
triangles corresponds to asymmetric ones. The SSB occurs in a certain
range of average density $\rho^*_{ssb1} \le \rho^*_{av} \le
\rho^*_{ssb2}$.}
\end{figure}

For temperatures below $T_w=110.1$~K, we obtained large ranges of
$\rho^*_{av}$ where the asymmetric solutions exhibit a lower free energy
than the corresponding symmetric ones. In spite of the fact that there
is a general idea that a connection exists between the SSB effect and 
nonwetting, we have found, by contrast, that SSB behavior extends above
the wetting temperature. Furthermore, we have also found a relation between
prewetting and SSB.

\begin{figure}[th]
\centering\includegraphics[width=0.48\textwidth]{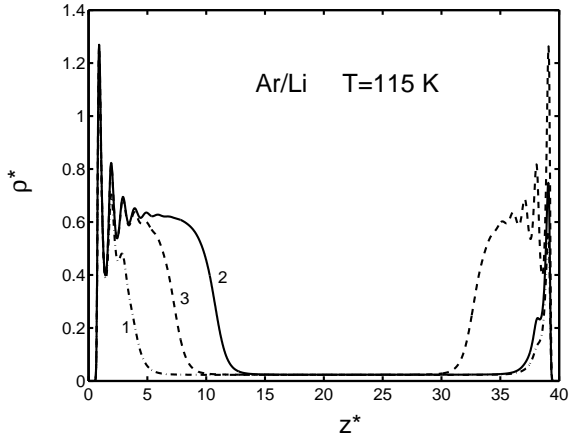}
\caption{\label{fig:profiles} Density profiles of Ar confined in a
slit of Li with $L^*=40$ at $T=115$~K. The displayed spectra denoted by
1, 2 and 3 correspond to average densities $\rho^*_{av}=0.074,
0.192$ and $0.218$, respectively.}
\end{figure}

Figure \ref{fig:free_en} shows the free energy per particle, $f_{\rm DF}
=F_{\rm DF}/N$, for both symmetric and asymmetric solutions for the
Ar/Li system at $T=115$~K$\,>T_w$ as a function of the average density.
According to this picture, the ground state (g-s) exhibits asymmetric
profiles between a lower and an upper limit $\rho^*_{ssb1}=0.057 \le
\rho^*_{av} \le \rho^*_{ssb2}=0.192$. Out of this range no asymmetric
solutions were obtained form the set of Eqs.\ (\ref{euler1})-(\ref{lambda}). Similar features were obtained for higher temperatures
until $T=118$~K, above this value the profiles corresponding to the g-s
are always symmetric. Figure \ref{fig:profiles} shows three examples of
solutions determined at $T=115$~K. The result labeled 1 is a small
asymmetric profile, that labeled 2 is the largest asymmetric
solution at this temperature. So, by further increasing $\rho^*_{av}$,
the SSB effect disappears and the g-s becomes symmetric, as indicated by
the curve labeled 3. When the asymmetric profiles occur, the situation
is denoted as partial (or one wall) wetting. The symmetric solutions
account for a complete (two wall) wetting. These different situations
can be interpreted in terms of the balance of $\gamma_{sl}$,
$\gamma_{sv}$ and $\gamma_{lv}$ surface tensions, carefully discussed
in previous works \cite{sikkenk87,nijmeijer90,szybisz08}. Here we shall
restrict ourselves to briefly outline the main features. When the
liquid is adsorbed symmetrically like in the case of profile 3 in Fig.\
\ref{fig:profiles}, there are two $s$-$l$ and two $l$-$v$ interfaces.
Hence, the total surface excess energy may be written as

\begin{equation}
\gamma^{sym}_{tot} = 2\,\gamma_{sl} + 2\,\gamma_{lv} \;.
\label{symm_p}
\end{equation}
On the other hand, for a asymmetric profile $\gamma^{asy}_{tot}$ becomes

\begin{equation}
\gamma^{asy}_{tot} = \gamma_{sl} + \gamma_{lv} + \gamma_{sv} \;.
\label{asym_p}
\end{equation}
The three quantities of the r.h.s. of this equation are related by Young's law (see, e.g., Eq.\ (2.1) in Ref.\ \cite{deGennes85})

\begin{equation}
\gamma_{sv} = \gamma_{sl} +\gamma_{lv}\,\cos{\theta} \:, \label{young}
\end{equation}
where $\theta$ is the contact angle defined as the angle between the
wall and the interface between the liquid and the vapor (see Fig. 1 in
Ref.\ \cite{deGennes85}). By using Young's law, the Eq.\
(\ref{asym_p}) may be rewritten as

\begin{equation}
\gamma^{asy}_{tot} = 2\,\gamma_{sl} + \gamma_{lv}\,(1 + \cos{\theta})
\;, \label{phase}
\end{equation}
with $\cos{\theta}=(\gamma_{sv}-\gamma_{sl})/\gamma_{lv}<1$. If one
changes $\gamma_{sl}$ by increasing enough $N_s$ (as shown in Fig.\
\ref{fig:free_en}), and/or $T$, and/or the strength of $U_{sf}(z)$,
eventually the equality $\gamma_{sv} - \gamma_{sl}=\gamma_{lv}$ may be
reached yielding $\cos{\theta}=1$. Then, the system would undergo a
transition to a symmetric profile where both walls of the slit are wet.

It is important to remark that, indeed, there are two degenerate
asymmetric solutions. Besides that one shown in Fig.\ \ref{fig:profiles}
where the profiles exhibit the thicker film adsorbed on the left wall
(left asymmetric solutions - LAS), there is an asymmetric solution with
exactly the same free energy but where the thicker film is located
near the right wall (right asymmetric solutions - RAS).

The asymmetry of density profiles may be measured by the quantity

\begin{equation}
\Delta_N = \frac{1}{N_s} \int^{L/2}_0 dz\,[\rho(z) - \rho(L-z)] \;.
\label{asymm}
\end{equation}
According to this definition, if the profile is completely asymmetrical
about the middle of the slit, i.e. for: (i) $\rho(z < L/2) \ne 0$ and
$\rho(z \ge L/2)=0$; or (ii) $\rho(z < L/2)=0$ and $\rho(z \ge L/2)
\ne 0$ this quantity becomes $+1$ or $-1$, respectively, while for
symmetric solutions it vanishes.
 
\begin{figure}[th]
\centering\includegraphics[width=0.48\textwidth]{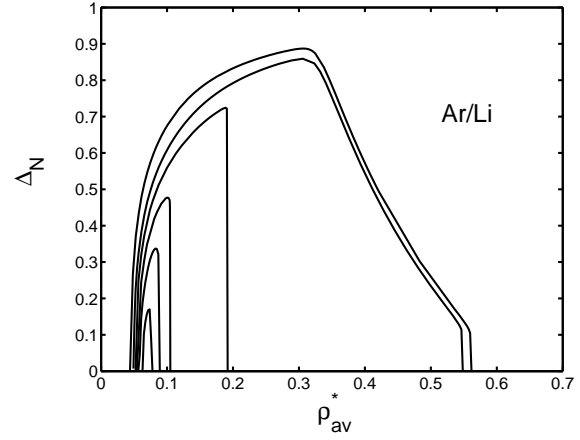}
\caption{\label{fig:asymmetry} Asymmetry parameter for Ar confined by
two Li walls separated by a distance of $L^*=40$ as a function of
average density. From outside to inside the curves correspond to
temperatures $T=112, 114, 115, 116, 117$ and 118~K. The asymmetric
solutions occur for different ranges $\rho^*_{ssb1} \le \rho^*_{av} \le
\rho^*_{ssb2}$.}
\end{figure}
 
\begin{figure}[th]
\centering\includegraphics[width=0.48\textwidth]{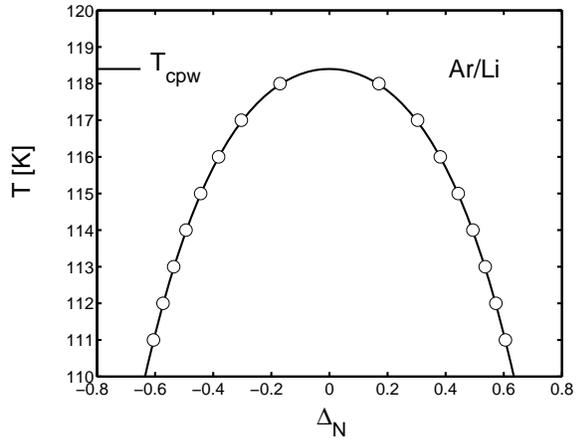}
\caption{\label{fig:ArLi_Tcpw} Circles stand for both branches of the
asymmetry parameter for Ar confined in an $L^*=40$ slit of Li walls for
temperatures between $T_w$ and $T_{cpw}$. The solid curve is the fit to
Eq.\ (\ref{Tcpw}) used to determine $T_{cpw}$.}
\end{figure}

We evaluated the asymmetry coefficients of solutions obtained for
increasing temperatures up to $T=118$~K. The results for LAS profiles
at temperatures larger that $T_w$ are displayed in Fig.\
\ref{fig:asymmetry} as a function of the average density. One may
observe how the range $\rho^*_{ssb1} \le \rho^*_{av} \le \rho^*_{ssb2}$
diminishes under increasing temperatures. The SSB effect persists at
most for the critical $\rho^*_{av}(crit)=(17/24)\,{\tilde\sigma}^2_{ff}
\times 10^{-2} \simeq 0.074$ with ${\tilde\sigma}_{ff}$ expressed
in \AA.

We shall demonstrate that by analyzing the data of $\Delta_N$ for
$\rho^*_{av}(crit)$ it is possible to determine the critical prewetting
point. Figure \ref{fig:ArLi_Tcpw} shows these values for both the
LAS and RAS profiles, calculated at different temperatures, suggesting a
rather parabolic shape. So, we propose a fit to the following quartic
polynomial

\begin{equation}
T = T_{cpw} + a_2 \Delta^2_N + a_4 \Delta^4_N \;. \label{Tcpw}
\end{equation}
This procedure yielded $T_{cpw}=118.4$~K, $a_2=-14.14$~K, and $a_4 =
-16.63$~K. The obtained value of $T_{cpw}$ is in agreement with the
limits established when analyzing the adsorption isotherms of the
one-wall systems displayed in Fig.\ \ref{fig:ArLi_Dmu}. These results
indicate that the disappearance of the SSB effect coincides with the
end of the prewetting line.

\section{Conclusions}
\label{sec:conclusions}

We have performed a consistent study within the same DF approach of free
slabs of Ar, the adsorption of these atoms on a single planar wall of
Li and its confinement in slits of this alkali metal. Good results
were obtained for the surface tension of the liquid-vapor interface.
The analysis of the physisorption on a planar surface indicates that Ar
wets surfaces of Li in agreement with previous investigations. The
isotherms for the adsorption on one planar wall exhibit a locus of
prewetting in the $\mu-T$ plane. A fit of such data yielded a wetting
temperature $T_w=110.1$~K. In addition, these isotherms also show that
the critical prewetting point $T_{cpw}$ lies between $T=118$ and 119~K.
These results for $T_w$ and $T_{cpw}$ are slightly below the values
obtained in Refs.\ \cite{ancilotto99,curtarolo00}, the discrepancy is
discussed in the text.

On the other hand, this investigation shows that the profiles of Ar
confined in a slit of Li present SSB. This effect occurs in a certain
range of average densities $\rho^*_{ssb1} \le \rho^*_{av} \le
\rho^*_{ssb2}$, which diminishes for increasing temperatures. The
main output of this work is the finding that above the wetting
temperature the SSB occurs until $T_{cpw}$ is reached. To the best of
our knowledge this is the first time that such a correlation is
reported. Furthermore, it is shown that by examining the evolution of
the asymmetry coefficient one can precisely determine $T_{cpw}$. The
obtained value $T_{cpw}=118.4$~K lies in the interval established when
analyzing the adsorption on a single wall.

\begin{acknowledgements}
This work was supported in part by the Grants PICT 31980/5 from Agencia
Nacional de Promoci\'on Cient\'{\i}fica y Tecnol\'ogica, and X099 from
Universidad de Buenos Aires, Argentina.
\end{acknowledgements}

\end{document}